\documentclass[prl,twocolumn,a4paper,showkeys]{revtex4}
\usepackage{graphicx}

\begin{document}

\newcommand{\be}{\begin{equation}}
\newcommand{\ee}{\end{equation}}
\newcommand{\bq}{\begin{eqnarray}}
\newcommand{\eq}{\end{eqnarray}}
\newcommand{\bc}{\begin{center}}
\newcommand{\ec}{\end{center}}
\newcommand{\pd}{\partial}
\newcommand{\Mpc}{{\mathrm{Mpc}}}

\title{Powering AGNs with super-critical black holes}

\author{Anastasios Avgoustidis\footnote{E-mail: tasos@ieec.uab.es}$^{a}$, Raul Jimenez\footnote{E-mail: raul@ieec.uab.es}$^{a,b}$, Luis \'Alvarez-Gaum\'e\footnote{E-mail: Luis.Alvarez-Gaume@cern.ch}$^c$, Miguel A. V\'azquez-Mozo\footnote{E-mail: Miguel.Vazquez-Mozo@cern.ch}$^d$}
\affiliation{$^a$Institute of Space Sciences (IEEC-CSIC), Faculty of Sciences, Campus UAB, 08193 Bellaterra, Spain\\
$^b$ ICREA \\
$^c$Theory Group, Physics Department, CERN,\\ CH-1211 Geneva 23, Switzerland\\
$^d$Dep.\ de F\'isica Fundamental, Univ.\ de Salamanca,\\ Pl.\ de la
Merced s/n, E-37008 Salamanca, Spain}

\date{\today}

\begin{abstract}
We propose a novel mechanism for powering the central engines of Active Galactic Nuclei
through super-critical (type II) black hole collapse.  In this picture, $\sim 10^3\
M_\odot$ of material collapsing at relativistic speeds can trigger a gravitational
shock, which can eject a large percentage of the collapsing matter at relativistic speeds,
leaving behind a ``light" black hole.  In the presence of a poloidal magnetic field, the
plasma collimates along two jets, and the associated electron synchrotron radiation can
easily account for the observed radio luminosities, sizes and durations of  AGN jets.
For Lorentz factors of order 100 and magnetic fields of a few hundred $\mu G$,
synchrotron electrons can shine for $10^6\ \mbox{\it yrs}$, producing jets of sizes of order
$100\ \mbox{\it kpc}$.  This mechanism may also be relevant for Gamma Ray Bursts and,
in the absence of magnetic field, supernova explosions.
\end{abstract}


\maketitle

\paragraph{Introduction -}

Black Holes (BH) are omnipresent in many astrophysical environments ranging from the centers of galaxies to the end products of massive stars. Their mass range is equally large: from a few to $\sim 10^9$ M$_{\odot}$. They are believed to play a central role in many astrophysical processes like the radio jets observed in Active Galactic Nuclei (AGN), the central engines of Gamma Ray Bursts (GRB), etc. Yet, the physical mechanism by which they influence their environment is still not firmly established. The most attractive proposal is the rotational energy of the BH by its interaction with the  surrounding magnetic field that in turn creates an electromotive force. This can be done via the Penrose \cite{Penrose} or, the most widely accepted, Blandford-Znajek mechanism \cite{BlanZnaj}. Because large ultra-relativistic factors are required to explain the jets in AGNs and GRBs, the extraction mechanism must be very efficient. However, theoretical models show difficulties in extracting the energy via this mechanism as it is not clear
how much energy can be extracted from the BH and accretion disk and whether or not it can achieve the required power to explain the observed astrophysical phenomena (see e.g. Ref.~\cite{Li}). Further, in the case of AGNs, it has always been problematic to feed the BH from a large accretion disk, as it is very difficult to envision how the material can be transported from very large distances
toward the horizon and how to keep a large disk in place without fragmenting \cite{Goodman03}.

However, recent numerical studies (see Ref.~\cite{Gundlach} for a review) have provided a new picture of BH collapse.
In short, there is a regime in which  a significant fraction of the collapsing matter can be directly ejected at ultra-relativistic speeds, leaving behind a  ``small" BH.  In this model, the astrophysical interaction of BHs
with the surrounding medium is provided by the collapse itself and thus it does not require any extra energy
extraction.
It is surprising that this regime, known in the relativity literature as super-critical type II
collapse, has not been seriously considered in astrophysical contexts, with the notable exceptions of
Refs. \cite{NieJed} (primordial BHs) and \cite{Novak},\cite{NoblChopt} (relativistic neutron star
collapse).  Yet, type II collapse can arise when the rest mass energy of the collapsing matter is
dynamically irrelevant, in particular when the velocities involved are relativistic, so it may well be
applicable in several situations in high-energy astrophysics.
In this letter we point out that this newly
proposed mechanism could explain many of the astrophysical phenomena related to BH activity, and in
particular we present a toy model for radio-jets in AGNs.

\paragraph{Critical Gravitational Collapse -}

The phenomenon of critical BH formation
was
discovered by Choptuik in a numerical
study of spherically symmetric collapse of a massless scalar field \cite{Choptuik}.
He considered one-parameter families of initial conditions parametrised by a number $p$, such
that a BH is formed for $p\gg 1$, whereas when $p\ll 1$ matter disperses to infinity
after collapse without leaving any remnant behind. Numerical simulation of such a system showed
the existence of a critical value of the parameter $p=p_*$ separating solutions that contain BHs
in the asymptotic future from those that do not.

Choptuik found strong
numerical evidence that, by fine-tuning the initial data to the critical value within each
family, one could produce a BH of arbitrarily small mass that retains a fraction
of the initial collapsing matter, the rest being dispersed to infinity. The total amount of matter
that escapes depends on the initial conditions.

The numerical study
showed that
the mass of the formed BH obeys a \emph{scaling relation}:
\be\label{mass_scaling}
M_{\rm BH} \simeq c_f |p-p_*|^\delta \,
\ee
where the coefficient $c_f$ depends on the chosen parametrisation (and so does
the critical value $p_*$), while the exponent $\delta$ has the universal value
$\delta\simeq 0.374$ for all families of initial conditions.

Similar studies were subsequently carried out for different types of matter coupled
to gravity,
mostly within spherical symmetry, although
the case of axisymmetric gravitational collapse has also been explored
\cite{Gundlach}.
They revealed
that the
critical exponent $\delta$ depends on the type of matter considered and  the dimensionality of
spacetime~\cite{higher-D,A-GGVTV-M}. For example, for the radiation fluid ~\cite{EvCol,A-GGVTV-M}
the critical exponent in four dimensions is $\delta\simeq 0.356$.

It was also realized that, unlike the case of the massless scalar field, the size of
the BH at criticality can also be finite, i.e. there is a gap in the mass of the BH
at $p=p_{*}$.
This is referred to as type I criticality,
while those cases exhibiting power-law mass scaling are known as type II.
On general grounds type II criticality arises whenever the collapse problem has no intrinsic
mass scale, or this is dynamically irrelevant. This is the case, for example, for massive matter
collapsing at relativistic speeds.

Indeed, Novak \cite{Novak} (see also \cite{NoblChopt})
has studied the case of a collapsing neutron star, and found that for relativistic inward
velocities a shock develops, which can eject part of the infalling matter, leaving only
a small fraction of the initial mass to form a BH.  For sufficiently high
velocities, BHs of arbitrary small mass can be formed, and this mass obeys the
scaling relation (\ref{mass_scaling}), where the role of the scaling parameter $p$ is
played by the initial inward velocity.  This example demonstrates that type II phenomena
could be relevant in high-energy astrophysics.

\paragraph{Astrophysics: a toy model -}

In this letter we propose to apply the above ideas in a number of high-energy
astrophysical scenarios.  In particular, we point out that near-critical type II
collapse could provide a powerful engine for AGN jets.  The key point is that in
this case (but also in other astrophysical scenarios) the velocities involved
are ultra-relativistic so that particle rest-masses become irrelevant: the
relativistic equations have effectively no mass-scale and type II phenomena
can arise.

Consider a toy-model in which a spherical shell of matter is collapsing with
relativistic inward velocity.  As discussed in the previous section, a
similar situation has been studied by Novak \cite{Novak} in the case of
a relativistically collapsing neutron star.  For high enough velocities,
a BH of arbitrary small mass can be formed,
so any fraction of the collapsing mass can be ejected to infinity.

\begin{figure}
   \includegraphics[width=\columnwidth]{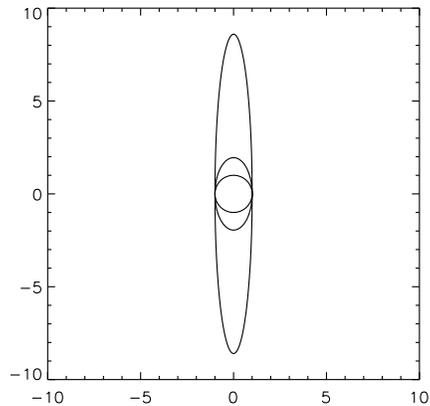}
   \caption{\label{spher_pol} Expanding shells of ionised matter in the
   presence of a magnetic field aligned with the y-axis. The shells
   start off spherical, but become collimated with time, as the charged
   particles' velocity component along the x-axis only produces circular
   motion of negligible radius.}
\end{figure}

Now consider a situation in which a large fraction of the infalling mass is being
ejected spherically outwards, but in the presence of a uniform magnetic field $\bf B$.
Charged particles with velocities $v$ orthogonal to $\bf B$ simply enter circular
orbits with radius $R=\gamma m v/q B$, frequency $\omega_0=Bq/\gamma m c$ ($m$, $q$ being
the particle mass and electric charge respectively), while particles moving along $\bf B$
feel no magnetic force.  A particle with pitch angle $\theta$ spirals along the magnetic
field lines, with radius proportional to $v \sin\theta$ and longitudinal speed equal
to $v \cos\theta$.  Assuming the radii of these helicoidal orbits are of astronomical
-- rather than cosmological -- size for the whole range of pitch angles\footnote{For $B
\simeq 1\,\mu G$ and $\gamma\simeq 10^5$, the maximum radius is $\sim 1\,AU$.  The radius
gets smaller for smaller $\gamma$ and larger $B$.}, this situation leads to a smooth transition
from spherical to poloidal geometry (Fig. \ref{spher_pol}).  The plasma gets collimated along
two outgoing jets, aligned with the magnetic field.  The energy available to the jets scales
with the ejected mass $M_{\rm ej}$, which, depending on the value of the initial inward
velocities, can be a sizeable fraction of the total mass available.

The late-time configuration produced in the above scenario strikingly resembles the
observed morphology of AGNs \cite{Krolik}.  The physics,
however, producing this picture is very different
from the one in the standard
AGN model. There, the energy needed to power the jets is extracted from the rotation of the
central BH through some variant of the Blandford-Znajek mechanism \cite{BlanZnaj}.
There has been significant concern whether this mechanism can adequately provide
the vast power seen in the jets (see for example~\cite{Li,komi}).
{ Also, the accretion disc -- more precisely the observed broad line region
believed to be associated to it -- is located several thousands of gravitational radii
away from the central BH, so it is difficult to connect accretion disc physics at
large distances to BH physics at the ergosphere region.}
In the picture of super-critical
collapse, which we are proposing, the efficiency can be much higher, as, in the
near-critical limit, a large fraction of the collapsing mass can be ejected at
relativistic speeds. This provides a powerful engine for feeding AGN jets.
{ Further, the energy comes from the collapse itself, so the jet is not directly
fed by the distant accretion disc.}

Another important difference to the standard picture is the way in which the particles
acquire their observed ultra-relativistic velocities.  In the standard scenario, the
plasma gets collimated to form two jets in a similar fashion (that is, through a
poloidal magnetic field), but the initial velocities are not relativistic, so one
also needs a mechanism to boost particles to large Lorentz factors, of order
$10$ or more.  Here, the shock developed in the final stages of the collapse ejects
the plasma at ultra-relativistic velocities to start with, so the only function of
the magnetic field is to collimate particle trajectories, rather than the more
difficult task of providing acceleration.  Thus, even in this simple toy-model where
magnetic gradient forces
are ignored, the observed large Lorentz factors can easily arise.

Given that the critical solution has zero measure in the space of initial conditions,
the obvious question is how close to criticality one would need to be in order to
provide sufficient power for the jet, and how ``unlikely" are the initial conditions
that would lead to this situation.  Then, one would have to explain how these initial
conditions could physically arise in an astrophysical set-up.  In the toy-model we
have considered, the initial conditions simply correspond to the initial inward velocity,
so the question reduces to finding a physical mechanism that would lead to ``fast enough"
collapse.  In the next section, we briefly examine these questions.

\paragraph{Is the model plausible? -}

Let us estimate the amount of mass needed to be ejected by the shock.  Each particle
in the plasma is following a helicoidal path with axis aligned with the magnetic field,
so it will emit synchrotron radiation as in the standard AGN model.  The total power
emitted by a single charged particle is:
\be\label{powerpp}
P_i = \frac{2}{3} \frac{q^2}{R} \omega_0 \left(\frac{v_\perp}{c}\right)^3 \gamma^4
= \frac{2}{3} \frac{q^4 B^2}{c^3 m^2} \left(\frac{v_\perp}{c}\right)^2 \gamma^2 \,,
\ee
where $v_\perp=v \sin\theta$ is the velocity component transverse to the magnetic field,
$\gamma$ is the Lorentz factor corresponding to $v$, and $c$ is the speed of light in vacuum.
Assuming that the plasma consists of approximately equal amounts of protons and electrons,
and noting that this power scales with the inverse square of the particle mass,
the total power of the jet (that is, summing over all particles) will be dominated by
the electron signal, so one should set $m=m_e$ for the particle mass in equation (\ref{powerpp}).

The total power in a typical radio-loud AGN jet\footnote{Note this is the total power
of the \emph{jet} only, which is practically the radio power.  The total \emph{AGN}
power is dominated by optical emission coming from the disc.} is of order $P_{\rm jet}
\sim 10^{43} \, \mbox{\it erg} \, s^{-1}$ (see for example~\cite{LiuJiangGu}), so for an AGN with
central magnetic field $B\sim 300\, \mu G$ one needs $P_{\rm jet}/P_i\sim 10^{60}$ electrons
with Lorentz factors of around $300$ to produce the observed luminosity from synchrotron
emission.  Then, there must be an approximately equal amount of protons of mass $m_p$
(which dominate the total mass) so the required ejected mass would be of the order of
$10^3\ M_\odot$.  Thus, for typical radio-loud AGNs, there is no need for fine
tuning to the near-critical limit; super-critical collapse of $10^3\, M_\odot$ where,
for example, $90\%$ of the infalling mass is ejected, leaving behind a $100\, M_\odot$
BH, can provide enough energy to explain observed luminosities.  Note, however,
that implicit to this is the assumption that the material is ejected at relativistic
speeds.  Since, the ejecta speed depends on both the equation of state and the initial
conditions \cite{NeilsChopt,NoblChopt}, one expects that this requirement will give
rise to some fine-tuning.  We are planning to quantitatively investigate this issue
in a follow-up publication \cite{AJA-GV}.

An obvious question is whether a synchrotron electron can shine for long enough to
explain the observed jet luminosities before it radiates away all its relativistic
energy.  For the magnetic field and Lorentz factors consider above, the synchrotron
lifetime of electrons is $\tau=\gamma m_e c^2 / P_i \sim 10^6\, \mbox{\it yrs}$, which is
consistent with AGN episodic lifetimes\cite{Tytler}.  The distance along the
magnetic field covered by the ultra-relativistic particles during this time
interval is $d\sim 300\, \mbox{\it kpc}$, which is again consistent with observed jet sizes.
The observed morphology of radio-loud AGNs \cite{Krolik} with a bright blob at the edge
of the jet fits very well in the above picture, where the ``bright'', radio emitting,
ultra-relativistic electrons make it first to the edge of the jet (producing the
luminous blob), while the rest, having a smooth distribution of speeds ranging from
$0$ to $c$ (the longitudinal speed is $v_{||}\simeq c\,\cos\theta$)
form the remaining of the jet structure.

Therefore our envisioned picture for how an AGN is powered is one in which a cloud of gas or
a star cluster of $\sim 10^3$ M$_{\odot}$  collapses into a near-critical BH. This can keep
taking place during the lifetime of the galaxy in several episodes. Here, it is worth noting
that the super-critical BH will typically be about one part in $10^5$ of the total mass of
the galaxy central BH.

\paragraph{Discussion -}

The scenario proposed here does not rely on the existence of an accretion disk whose angular
momentum fuels the jet. The key point is instead the formation of BHs by near-critical
gravitational collapse in which a fraction of the collapsing matter is dispersed back at
relativistic speeds\footnote{The accretion disk, however, might play an instrumental role in
generating the magnetic field responsible for the collimation of the ionised ejecta.}.
Therefore, if this mechanism is correct, there should be no strong correlation between
jet power and the rotational properties of the accretion disc.
This seems to be supported by
the fact that the optical and UV properties of radio-loud and quiet objects
are quite similar, despite very large differences in radio luminosities \cite{A_W&C}.
This also fits with the fact that most AGNs are found in elliptical galaxies, which are
gas-poor.

Another interesting point is that, in this picture, smaller jets should be related to BH
collapse further away from criticality, where a smaller fraction of collapsing material
is ejected.  In this situation one expects lower speeds of the ejecta, so smaller jets
should be associated with smaller Lorentz factors.

The above ideas could also apply in other similar contexts:  GRBs are short duration events,
in which the ejecta have very high Lorentz factors.  This could be explained by near-critical
BH formation at the center of the star, ejecting material that later interacts with the
stellar envelope.
The model is probably most natural for SN explosions, where the collapse is spherical
and certainly involves relativistic speeds.  In this case, however, there is no strong
magnetic field, so the final configuration would be closer to spherical rather than
poloidal.

Here we have presented a promising idea: near-critical type II collapse, a purely gravitational
mechanism that can eject matter at relativistic speeds, could, in the presence of a strong
magnetic field, be responsible for astrophysical jet formation.  We have demonstrated that
the collapse of $10^3\ M_\odot$ in a magnetic field of ${\cal O}(100)\ \mu G$ can easily account
for the observed radio luminosities of loud AGNs through electron synctrotron emission.
Although here we have discussed spherically symmetric collapse, the fact that the addition
of angular momentum does not spoil critical behavior \cite{angular_momentum} serves as
evidence that the proposed mechanism can be applied to realistic astrophysical situations.

There are many more directions to explore: it would be interesting to run perfect fluid collapse
simulations and check how easy it is to set up reasonable initial conditions that evolve to
sufficiently near-critical collapse.  One could also generalise the more formal analysis of
Ref.~\cite{A-GGVTV-M}, to include magnetic field and/or rotation.  Another interesting situation
would be to study the axisymmetric case, considering for example a disk collapsing to form a BH.
On the more astrophysical side, one could consider the specific application of this mechanism in
other astrophysical situations, e.g. GRBs and SN explosions, as mentioned above.

\paragraph{Acknowledgements -}
AA is supported by the Spanish National Research Council (CSIC) and FP7-PEOPLE-2007-4-3
IRG.  RJ is supported by FP7-PEOPLE-2007-4-3 IRG and by MICINN grant AYA2008-03531.
We would like to thank the University of Barcelona and in particular the Institute of
Cosmos Sciences for hospitality. M.A.V.-M. acknowledges partial support from Spanish Government
Grant FIS2006-05319, Basque Government Grant IT-357-07 and Spanish
Consolider-Ingenio 2010 Program CPAN (CSD2007-00042).

\end{document}